# Novel Quaternary Dilute Magnetic Semiconductor (Ga,Mn)(Bi,As): Magnetic and Magneto-Transport Investigations

K. Levchenko[1], T. Andrearczyk[1], J. Z. Domagala[1], J. Sadowski[1,2], L. Kowalczyk[1], M. Szot[1], R. Kuna[1], T. Figielski[1] and T. Wosinski[1]

[1]*Institute of Physics, Polish Academy of Sciences, 02-668 Warsaw, Poland*
[2]*MAX-IV Laboratory, Lund University, P.O. Box 118, SE-221 00 Lund, Sweden*

Magnetic and magneto-transport properties of thin layers of the (Ga,Mn)(Bi,As) quaternary dilute magnetic semiconductor grown by the low-temperature molecular-beam epitaxy technique on GaAs substrates have been investigated. Ferromagnetic Curie temperature and magneto-crystalline anisotropy of the layers have been examined by using magneto-optical Kerr effect magnetometry and low-temperature magneto-transport measurements. Post-growth annealing treatment of the layers has been shown to enhance the hole concentration and Curie temperature in the layers. Significant increase in the magnitude of magneto-transport effects caused by incorporation of a small amount of Bi into the (Ga,Mn)As layers, revealed in the planar Hall effect (PHE) measurements, is interpreted as a result of enhanced spin-orbit coupling in the (Ga,Mn)(Bi,As) layers. Two-state behaviour of the planar Hall resistance at zero magnetic field provides its usefulness for applications in nonvolatile memory devices.

___________________

Corresponding author:

K. Levchenko, Institute of Physics, Polish Academy of Sciences, Al. Lotnikow 32/46, 02-668 Warsaw, Poland, e-mail: levchenko@ifpan.edu.pl

## 1. Introduction

Last decades in the field of semiconductor physics were marked by tremendous improvement in the materials technologies, which might provide a basis for developing novel spintronic devices. Special attention was paid to the ternary III-V semiconductor (Ga,Mn)As, combining semiconducting properties with magnetism, which became a model among dilute magnetic semiconductors [1, 2]. Homogeneous layers of $Ga_{1-x}Mn_xAs$ containing up to above 10% of Mn atoms can be grown by a low-temperature (150–250°C) molecular-beam epitaxy (LT-MBE) [3, 4]. When intentionally undoped, the layers are of *p*-type where Mn atoms, substituting the Ga atoms in GaAs crystal lattice, supply both mobile holes and magnetic moments. Below the Curie temperature the layers become ferromagnetic due to the hole-mediated ordering of Mn spins. The sensitivity of their magnetic properties, such as the Curie temperature and magnetic anisotropy, to the growth-induced strain and hole concentration allows for tuning those properties by growth peculiarities or post-growth annealing of the (Ga,Mn)As layers. Moreover, appropriate nanostructurization of thin (Ga,Mn)As layers offers the prospect of taking advantage of magnetic domain walls in novel spintronic devices. In this context, our recent studies on several types of nanostructures patterned from ferromagnetic (Ga,Mn)As layers pointed to their utility for spintronic applications. Especially, nanostructures of the three-arm [5], cross-like [6] and ring-shape [7] geometries, displaying magneto-resistive effects controlled by manipulation of magnetic domain walls in the nanostructures, could be applied in a new class of nonvolatile memory cells.

On the other hand, the replacement of a small fraction of As atoms by much heavier Bi atoms in GaAs results, due to an interaction of Bi 6*p* bonding orbitals with the GaAs valence band maximum, in a rapid decrease in its band-gap energy [8, 9] and in a strong enhancement of spin-orbit coupling accompanied by a giant separation of the spin-split-off hole band in Ga(Bi,As) [10]. The enhanced spin-orbit coupling could be advantageous for spintronic materials as it strongly affects their magneto-transport properties. In order to explore this issue we have investigated an impact of Bi incorporation into (Ga,Mn)As layers on their structural, magnetic and magneto-transport properties. First homogeneous layers with a high structural perfection of the (Ga,Mn)(Bi,As) quaternary dilute magnetic semiconductor, containing up to 6% Mn and 1% Bi, have been recently grown with LT-MBE under compressive misfit strain, on GaAs substrate, [11, 12] as well as under tensile misfit strain, on (In,Ga)As buffer layer [13]. Their magnetic properties were similar to those of the ternary (Ga,Mn)As layers [14], with the in-plane and out-of-plane easy axis of magnetization in the layers grown under the compressive and tensile misfit strain, respectively [13]. Our results of magneto-transport characterization of those layers confirmed significant enhancement of their magnetoresistance and the planar Hall resistance as a result of Bi incorporation into (Ga,Mn)As layers [15]. In the present paper we report on magnetic and magneto-transport properties of thin epitaxial layers of (Ga,Mn)(Bi,As) quaternary compound grown under compressive misfit strain, proving their usefulness for possible spintronic applications.

## 2. Experimental

The investigated (Ga,Mn)(Bi,As) layers, of 10 nm thickness and 6% Mn and 1% Bi content, were grown on semi-insulating (001)-oriented GaAs substrate by the LT-MBE technique at a temperature of 230°C. For comparison, similar (Ga,Mn)As layers were grown under the same conditions. After the growth the layers were subjected to the annealing treatment in air at the temperature of 180°C during 50 h. Post-growth annealing of (Ga,Mn)As layers [16], as well as the (Ga,Mn)(Bi,As) ones [11], at temperatures below the growth temperature was proved to substantially improve magnetic and electrical-transport properties of the layers as a result of outdiffusion of charge- and moment-compensating Mn interstitials from the layers.

Raman spectroscopy was employed to estimate the hole densities in the *p*-type (Ga,Mn)As and (Ga,Mn)(Bi,As) layers. The micro-Raman measurements were performed using a confocal Raman microscope (MonoVista CRS+) at room temperature with the 532-nm semiconductor laser as an excitation source. Magnetic properties of the layers were examined using magneto-optical Kerr effect (MOKE) magnetometry. The MOKE experiments were performed in longitudinal geometry using He-Ne laser as a source of linearly polarized light with the laser spot of about 0.5 mm in diameter and the angle of light incidence on the sample of about 30°. The standard lock-in technique with photo-elastic modulator operating at 50 kHz and a Si diode detector was used. Measurements of magnetic hysteresis loops were performed in the temperature range $T = 6-150$ K and in external magnetic fields up to 2 kOe applied in the plane of the layer. Magneto-transport measurements were carried out on Hall bars of 100 µm width and 200 µm distance between the voltage contacts, aligned along the [$\bar{1}$10] crystallographic direction, prepared from the annealed (Ga,Mn)As and (Ga,Mn)(Bi,As) layers by means of electron-beam lithography patterning and chemical etching. Both the longitudinal resistance, $R_{xx}$, and the planar Hall resistance, $R_{xy}$, were measured simultaneously using low-frequency (18 Hz) lock-in technique.

The measurements were performed for various orientations of the in-plane magnetic field $H$ at the liquid-helium temperature range.

## 3. Results and discussion

The micro-Raman spectra for the (Ga,Mn)As and (Ga,Mn)(Bi,As) layers were recorded from the (001) surface in the backscattering configuration. From analysis of the spectra, especially the coupled plasmon–LO phonon (CPLP) mode present in the spectra of the Mn containing $p$-type layers [11, 17], we have estimated the hole concentrations of about $8.5 \times 10^{19}$ cm$^{-3}$ and $7 \times 10^{19}$ cm$^{-3}$ in the as-grown (Ga,Mn)As and (Ga,Mn)(Bi,As) layers, respectively. As a result of the annealing treatment the hole concentrations in the layers increased to about $1.6 \times 10^{20}$ cm$^{-3}$ and $1 \times 10^{20}$ cm$^{-3}$, respectively, in a quite good agreement with the results of our room-temperature Hall effect measurements.

The MOKE measurements for the annealed (Ga,Mn)As and (Ga,Mn)(Bi,As) layers performed as a function of magnetic field displayed nearly rectangular hysteresis loops evidencing for the in-plane magnetization. Analysis of the MOKE hysteresis loops obtained under a magnetic field along the main in-plane crystallographic directions indicated easy magnetization axes along the in-plane ⟨100⟩ cubic directions and hard axes along two magnetically nonequivalent in-plane ⟨110⟩ directions, with the [$\bar{1}$10] direction being magnetically easier than the perpendicular [110] one. Such a rather complicated magneto-crystalline anisotropy is characteristic of (Ga,Mn)As layers grown under compressive misfit strain [6, 14]. From the temperature dependences of MOKE magnetization hysteresis loops we have determined the ferromagnetic Curie temperatures, $T_C$, of about 87 K and 80 K for the (Ga,Mn)As and (Ga,Mn)(Bi,As) layers, respectively.

Electrical resistance in ferromagnetic materials depends on the angle $\theta$ between the directions of current magnetization vector, resulting in anisotropic magnetoresistance (AMR) and so-called planar Hall effect (PHE) [18]. PHE manifests itself as a spontaneous transverse voltage that develops, owing to the spin-orbit interaction, in response to longitudinal current flow under an in-plane magnetic field or even in the absence of an applied magnetic field. In single domain model the AMR components can be described by expressions [19]:

$$R_{xx} = R_\perp + (R_\parallel - R_\perp)\cos^2\theta \tag{1}$$

$$R_{xy} = \tfrac{1}{2}(R_\parallel - R_\perp)\sin 2\theta, \tag{2}$$

where $R_\perp$ and $R_\parallel$ are the resistances for in-plane magnetization vector oriented perpendicular and parallel to the current, respectively. The longitudinal MR is described by Eq. (1) and the transverse resistance, i.e. the planar Hall resistance, is described by Eq. (2). In ferromagnetic (Ga,Mn)As, in contradiction to metallic ferromagnets, $(R_\parallel - R_\perp) < 0$, i.e. the resistance is higher when the magnetization is perpendicular to the current with respect to that when they both are parallel [20]. The same also holds for (Ga,Mn)(Bi,As), as demonstrated in our previous paper [15]. Moreover, the magnitude of PHE in (Ga,Mn)As layers can be up to four orders of magnitude greater than previously found in metallic ferromagnets, which was called as the giant PHE [20]. Very large magnitude of PHE observed in (Ga,Mn)As results primarily from the combined effects of strong spin-orbit interaction in the valence band of the zinc blende crystal structure and the large spin polarization of holes in (Ga,Mn)As.

Normalized longitudinal MR measured for the Hall bars of (Ga,Mn)As and (Ga,Mn)(Bi,As) layers under magnetic field parallel to the current is presented in Fig. 1. The up-and-down magnetic field sweep in the range of ±1 kOe results in nonmonotonic behaviour at low fields, described by Eq. (1), superimposed on isotropic negative MR extended to higher fields. Negative MR is a common property of ferromagnetic metals resulting from the reduction of spin-disorder scattering of charge carriers due to alignment of magnetic-ion spins by an external magnetic field. However, in (Ga,Mn)As, at the lowest temperatures when the spins are fully ferromagnetically ordered, the negative MR results mainly from the magnetic-field-induced destruction of quantum interference contribution to the resistivity caused by the effect of weak localization [21].

Figure 2 presents the planar Hall resistance results for the Hall bars of the two layers measured under magnetic field perpendicular to the current. While sweeping the magnetic field in the range of ±1 kOe, $R_{xy}$ displays the double hysteresis loop behaviour with recurrence points at magnetic fields where the magnetization vector in the Hall bar is forced to be directed along the external magnetic field. The planar Hall resistance changes between two extremum values, which, according to Eq. (2), are obtained at the angles $\theta$ of 45°, 135°, 225° and 315° corresponding to the orientations of the magnetization vector along the in-plane ⟨100⟩ crystallographic directions. The up-and-down magnetic field sweep in the whole range causes a rotation of the magnetization vector by 360° between all the four in-plane ⟨100⟩ directions, which represent equivalent easy axes of magnetization defined by cubic magneto-crystalline anisotropy.

While sweeping a magnetic field in a narrow range of ±200 Oe, $R_{xy}$ displays a single hysteresis loop analogous to the magnetization hysteresis loop, as shown in Fig. 3. Here, the magnetic field scan results in 90° rotations of the magnetization vector between two in-plane ⟨100⟩ directions and the results are consistent with the expected single-domain structure of the Hall bar. This hysteretic behaviour of the planar Hall resistance, which can assume one of the two stable values at zero magnetic field, that depends on the previously applied field, may be applicable in memory elements. All the results presented in Figs. 1, 2 and 3 demonstrate much larger magnitudes of both the longitudinal MR changes and the PHE resistances revealed for the Hal bar of the (Ga,Mn)(Bi,As) layer, with respect to that of the (Ga,Mn)As one, resulting from an enhancement of the spin-orbit interaction upon the addition of Bi into the layers.

## 4. Conclusions

Homogeneous layers of the (Ga,Mn)(Bi,As) quaternary dilute magnetic semiconductor have been grown by the low-temperature MBE technique on GaAs substrate. A small amount of Bi incorporated into (Ga,Mn)As significantly enhances the strength of spin-orbit coupling in this material, which manifests itself in its magneto-transport properties as an increase in both the AMR changes and the PHE resistance, and makes this novel dilute magnetic semiconductor advantageous for spintronic applications. Moreover, we have demonstrated that the PHE can be very useful in the study of magnetic properties of ferromagnetic semiconductors and their structures. The PHE can be also used as a basis for a new type of a nonvolatile memory elements, in which a bit of information is written magnetically and read electrically.

**Acknowledgement** This work has been partly supported by the Polish National Science Centre under grant No. 2011/03/B/ST3/02457.

**FIGURE CAPTIONS**

Fig. 1. Relative longitudinal resistance $R_{xx}$ for the Hall bars of (Ga,Mn)As and (Ga,Mn)(Bi,As) layers measured at the temperature 4.2 K, while sweeping an in-plane magnetic field, parallel to the current, in opposite directions, as indicated by arrows. The curves have been vertically offset for clarity.

Fig. 2. Planar Hall resistance $R_{xy}$ for the Hall bars of (Ga,Mn)As and (Ga,Mn)(Bi,As) layers measured at the temperature 4.2 K, while sweeping an in-plane magnetic field, perpendicular to the current, in opposite directions, as indicated by arrows. The curves have been vertically offset for clarity.

Fig. 3. Same as in Fig. 2 but for narrower range of sweeping magnetic field.

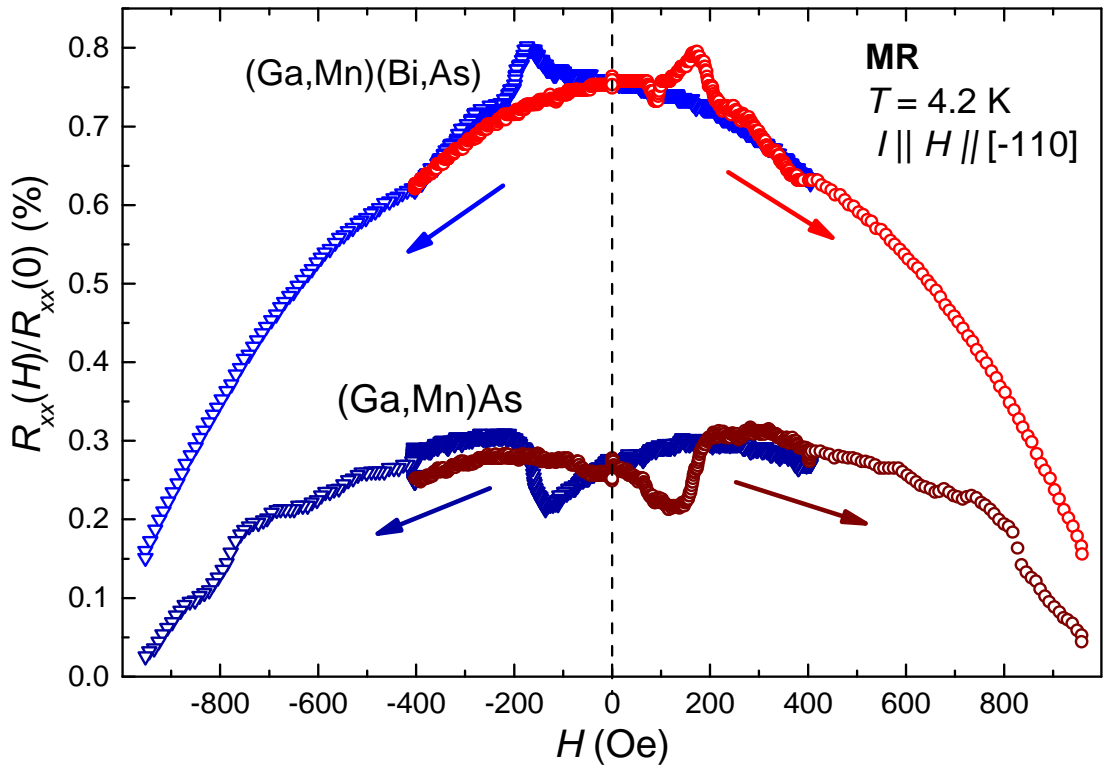

Fig. 1.

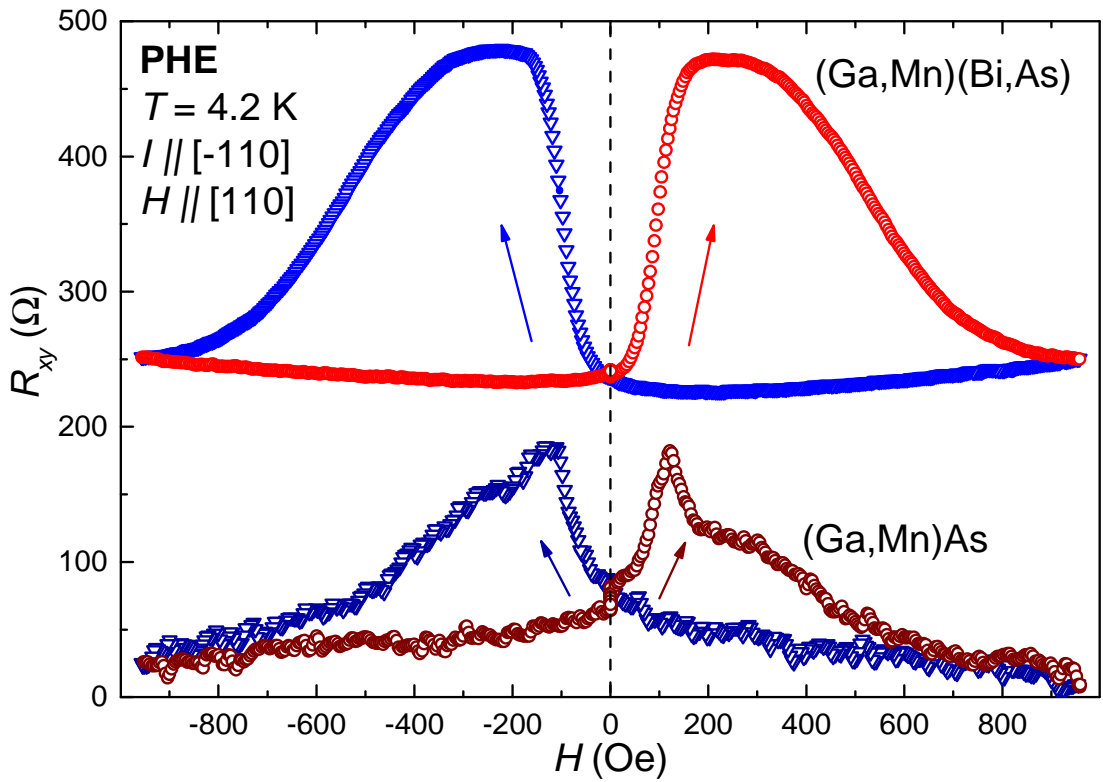

Fig. 2.

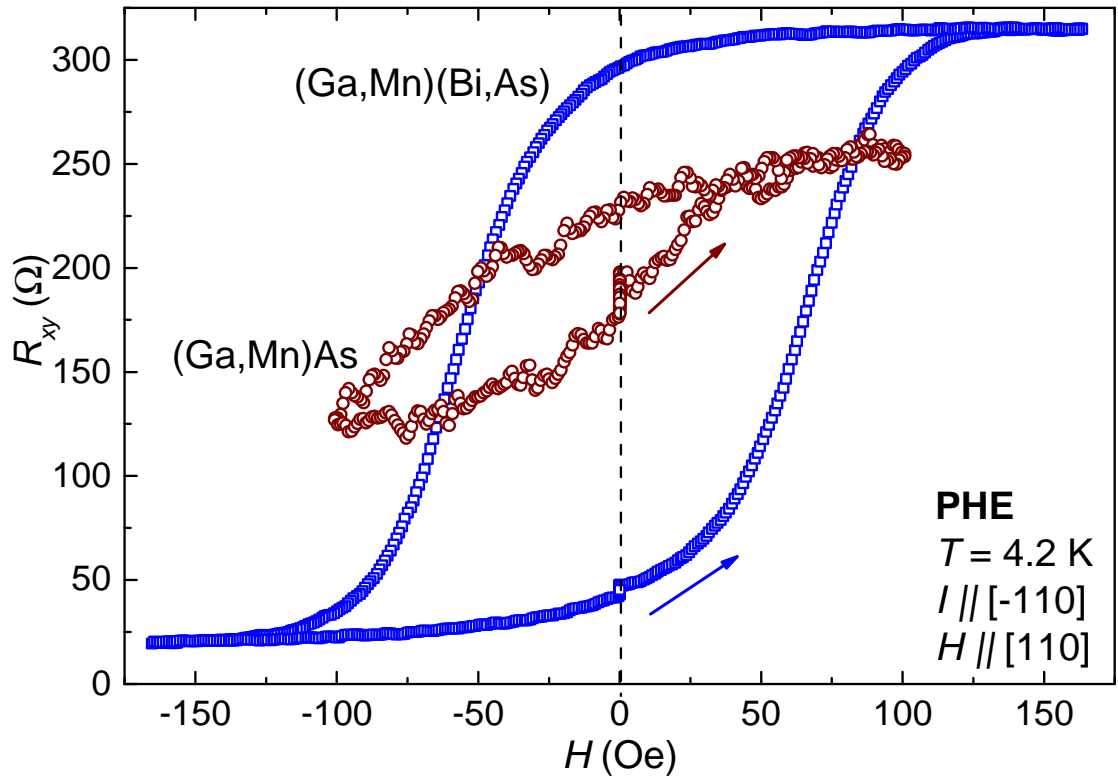

Fig. 3